\documentclass[aps,prl,reprint,superscriptaddress]{revtex4-2}
\usepackage{graphicx}
\usepackage{hyperref}

\begin{document}


\title{Controlling external injection in laser-plasma accelerators with terahertz frequency bunch manipulation}

\author{Aras Amini}
\email{aras.amini@manchester.ac.uk}
\affiliation{The Cockcroft Institute, Sci-Tech Daresbury, Keckwick Lane, Warrington WA4 4AD, UK}
\affiliation{Department of Physics and Astronomy \& Photon Science Institute, The University of Manchester, Oxford Road, Manchester M13 9PL, UK}

\author{Lewis R. Reid}
\affiliation{The Cockcroft Institute, Sci-Tech Daresbury, Keckwick Lane, Warrington WA4 4AD, UK}
\affiliation{Accelerator Science and Technology Centre, Science and Technology Facilities Council, Sci-Tech Daresbury, Keckwick Lane, Warrington WA4 4AD, UK}

\author{James K. Jones}
\affiliation{The Cockcroft Institute, Sci-Tech Daresbury, Keckwick Lane, Warrington WA4 4AD, UK}
\affiliation{Accelerator Science and Technology Centre, Science and Technology Facilities Council, Sci-Tech Daresbury, Keckwick Lane, Warrington WA4 4AD, UK}

\author{Morgan T. Hibberd}
\affiliation{The Cockcroft Institute, Sci-Tech Daresbury, Keckwick Lane, Warrington WA4 4AD, UK}
\affiliation{Department of Physics and Astronomy \& Photon Science Institute, The University of Manchester, Oxford Road, Manchester M13 9PL, UK}

\author{Laura Corner}
\affiliation{The Cockcroft Institute, Sci-Tech Daresbury, Keckwick Lane, Warrington WA4 4AD, UK}
\affiliation{School of Engineering, University of Liverpool, The Quadrangle, Brownlow Hill, Liverpool L69 3GH, UK}

\author{Darren M. Graham}
\affiliation{The Cockcroft Institute, Sci-Tech Daresbury, Keckwick Lane, Warrington WA4 4AD, UK}
\affiliation{Department of Physics and Astronomy \& Photon Science Institute, The University of Manchester, Oxford Road, Manchester M13 9PL, UK}

\author{Steven P. Jamison}
\affiliation{The Cockcroft Institute, Sci-Tech Daresbury, Keckwick Lane, Warrington WA4 4AD, UK}
\affiliation{Department of Physics, Lancaster University, Bailrigg, Lancaster LA1 4YB, UK}

\author{Graeme Burt}
\affiliation{The Cockcroft Institute, Sci-Tech Daresbury, Keckwick Lane, Warrington WA4 4AD, UK}
\affiliation{School of Engineering, Lancaster University, Bailrigg, Lancaster LA1 4YW, UK}

\author{Robert B. Appleby}
\affiliation{The Cockcroft Institute, Sci-Tech Daresbury, Keckwick Lane, Warrington WA4 4AD, UK}
\affiliation{Department of Physics and Astronomy \& Photon Science Institute, The University of Manchester, Oxford Road, Manchester M13 9PL, UK}

\date{\today}

\begin{abstract}
Laser-plasma wakefield acceleration (LWFA) offers ultrahigh accelerating gradients in compact setups, but the complex non-linear nature of the process makes it challenging to generate high-quality beams. Injection of electron bunches from an external source into a plasma accelerator provides a promising route to improved performance; however, electron bunches from conventional radio-frequency (RF)-based injectors suffer from non-linear compression and laser-beam asynchrony, leading to energy jitter and emittance growth. We present a fundamental concept of terahertz-controlled electron bunches for external injection into LWFA. This terahertz-frequency approach provides temporal locking between the electron beam and the drive laser, and enables the compression of high-quality beams to sub-10-fs durations before injection into the LWFA. Numerical simulations demonstrate that GeV-scale acceleration with excellent beam quality and stability---energy jitter and energy spread around 0.2\%---can be achieved using this method. This concept opens new opportunities for stable, multi-stage laser-driven accelerators and supports the development of next-generation applications such as free-electron lasers (FELs).

\end{abstract}

\maketitle


Laser-driven acceleration methods \cite{tajima,HOFI1,Hibberd2020,beamdr} offering a promising pathway toward compact and high-gradient acceleration. Among these, laser-plasma wakefield acceleration (LWFA) has gained significant attention \cite{PRl1,prl2,prl3,Leemans2006} for producing GeV-class electron beams. While gradients of several gigavolts per meter (GV/m) are attainable with LWFA, the complex non-linear interaction of the laser with the plasma imposes stringent challenges to generating high-quality bunches (i.e. with low-emittance and narrow energy spread). These are critical for applications in high-energy physics \cite{prlhep}, where high luminosity and collision rates demand tight beam quality control, and in free-electron lasers (FELs) \cite{fel}, where high-brightness coherent radiation typically requires per-mille level energy spread and normalized emittance below 1\,$\mu$m \cite{prl2fel,prl3fel,prl4fel}.

LWFA beams are predominantly generated through internal (or self-) injection, where the beam originates directly from the background plasma. These schemes rely on trapping plasma electrons through mechanisms such as ionization injection \cite{ion3,ion2,ion1}, shock-front injection \cite{shock1,shock2}, or more sophisticated variants \cite{sof1,soph3,sof2}. However, these schemes are highly sensitive to plasma conditions and nonlinear laser–plasma dynamics, making injection difficult to control \cite{OumbarekEspinos2023}. As a result, self-injection often suffers from poor shot-to-shot reproducibility, large charge fluctuations, and broadband energy spectra \cite{soph3,vGrafenstein2023}, posing significant challenges to generating stable, high-quality bunches. An alternative approach is external injection \cite{ex2,ex3,Wu2021}, which involves introducing electron bunches from an independent source into the plasma wakefield, allowing better control over the quality of these injected bunches. The strong correlation between injected and extracted bunch quality, together with the demand for multi-stage LWFA \cite{multi}, has driven global research efforts \cite{desyPhysRevAccelBeams.19.054401,ROSSI2014} toward the development of reliable methods for providing optimized external electron bunches for LWFA applications, frequently utilizing conventional radio frequency (RF) injectors.

Studies \cite{impact1-PhysRevSTAB.10.121301,eff} show that the injected electron bunch must be shorter than one-quarter of the plasma wavelength ($\lambda_p/4$) and synchronized with the laser-plasma wakefield with a precision of less than 10$\%$ of the plasma period ($T_p$). 
A nominal laser-plasma accelerator with a plasma density of 10$^{17}$\,cm$^{-3}$ has a plasma wavelength and plasma period of 105\,$\mu$m, and 352\,fs, respectively. 
Therefore, to achieve optimal injection into such an LWFA requires electron bunches with a duration shorter than 88\,fs and synchronized to the plasma wakefield to less than 35\,fs. Conventional RF injectors typically rely on a magnetic compression system to provide ultrashort electron bunches \cite{rf_lin,rf_lic2,jitter} and studies \cite{desyPhysRevAccelBeams.19.054401,Wu2021} highlight the significant challenges in producing ultrashort electron bunches suitable for LWFA applications. Achieving ultrashort bunches with high stability in practice remains an obstacle, primarily due to nonlinearities and jitter introduced by the RF cavities coming from their mutual independence and asynchronization with the drive laser. Alternative approaches for improving stability in LWFA-based facilities include optical synchronization used in FELs \cite{synch_Schulz2015}, machine learning and active feedback loops \cite{stab-PhysRevAccelBeams.22.041303,stab-Shalloo2020,stab-PhysRevLett.126.104801}, and post-plasma energy compression \cite{stab_main_PhysRevLett.129.094801,Winkler2025}. While these methods have shown varying degrees of success,  their experimental realization at GeV-level LWFA-based facilities remains an open challenge.

In this letter, we present a new method of generating sub-femtosecond electron bunches with precise synchronization for external injection into a plasma accelerator by using THz-driven waveguides. These waveguides have received considerable interest as they offer a promising route to femtosecond control of electron beams \cite{comp2,compre}, with our recent work \cite{Hibberd:2025kkc} demonstrating the ability to compress high-energy electron beams while achieving temporal locking and suppression of jitter. Here, we first develop a theoretical framework for a THz-driven compression system, emphasizing how higher-frequency fields and precise laser-based synchronization enable performance comparable to advanced RF-based systems. We then present start-to-end numerical modeling demonstrating energy stabilization and high-quality accelerated bunches. These stable bunches open a path toward multi-stage LWFA, with potential applications including direct injection into storage rings and plasma-based colliders.

\begin{figure*}
    \centering
    \includegraphics[width=1\textwidth]{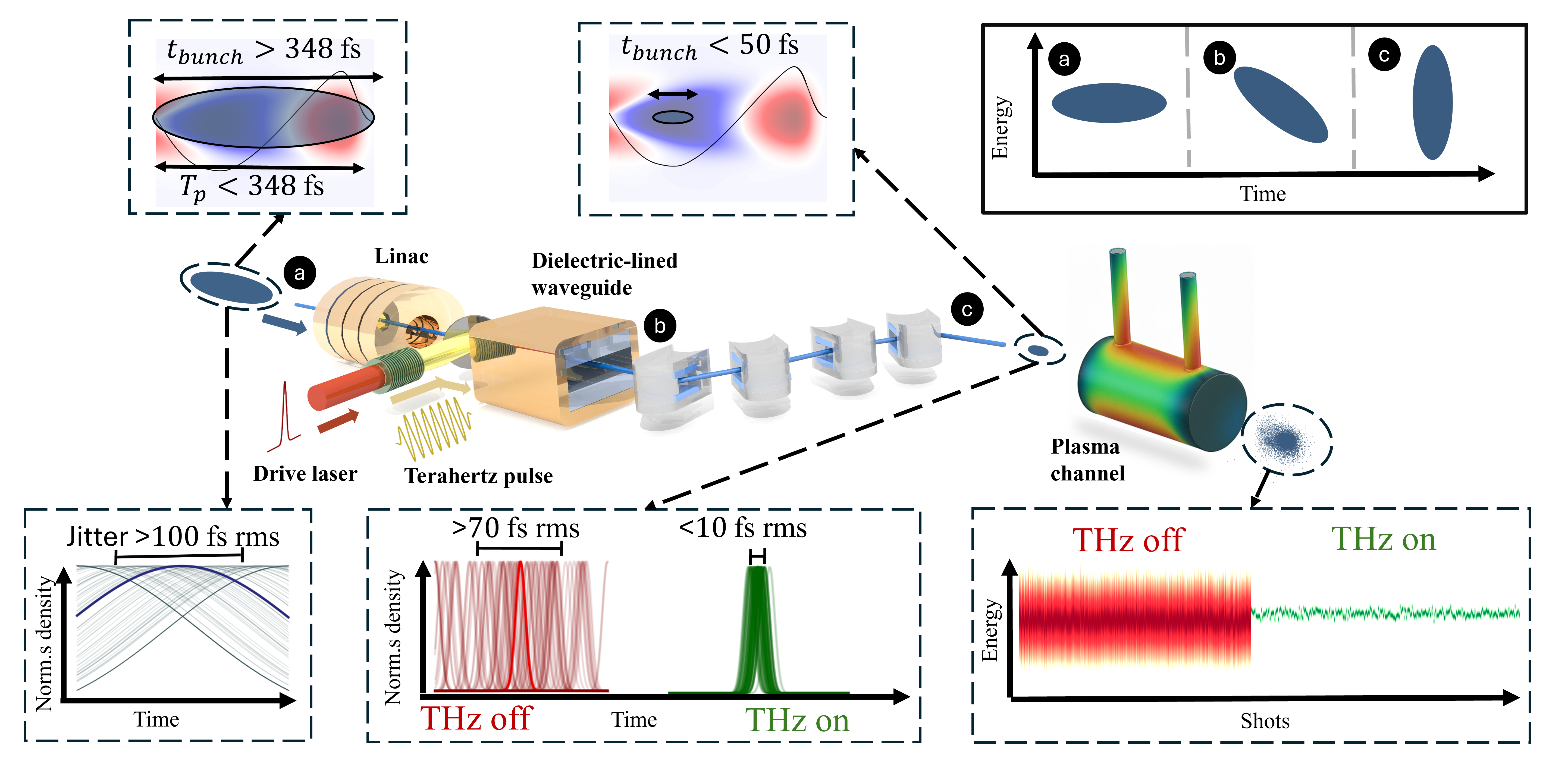} 
    \caption{A schematic diagram demonstrating THz compression and injection of an electron bunch into a plasma accelerator. (a) The initial long electron bunch requires compression (blue) relative to the plasma period ($T_p$). This compression is achieved by imparting a linear energy chirp on the bunch with a THz-driven waveguide (b), followed by a magnetic chicane (c) that performs longitudinal compression and matching into the LWFA plasma stage. While conventional compression methods produce short bunches, the final injected bunches (red) suffer from large arrival-time jitter relative to the plasma wakefield. In contrast, the proposed THz-controlled scheme (green) achieves the required longitudinal compression while simultaneously suppressing time jitter.}
    \label{fig:1}  
    \label{fig:1a} 
    \label{fig:1b} 
    \label{fig:1c} 
\end{figure*}

The concept of THz-controlled electron bunches for external injection into an LWFA takes advantage of the intrinsic synchronization between THz pulse generation and plasma wakefield excitation enabled by using a common laser system to drive both processes, enabling the delivery of stable, high-quality ultrashort bunches. The concept is illustrated schematically in Fig.\ref{fig:1a}, where electron bunches (a) from an external RF source (injector and linac) interact with laser-generated THz pulses inside a dielectric-lined waveguide (DLW). The THz-driven DLW (b) acts as a linear energy chirper, imprinting a time-energy correlation on the bunch that intentionally increases its energy spread. This is followed by a magnetic chicane for longitudinal phase-space rotation, compression, and transverse matching into the final stage of laser-plasma acceleration (c).

With respect to a reference bunch fully synchronized with the laser–plasma accelerator, this THz-controlled concept reduces the shot-to-shot arrival-time jitter of the electron bunch prior to LWFA compared to cases where THz pulse generation is off and the time–energy correlation is induced by RF linacs. This improvement arises from the intrinsic synchronization—similar to optical pump–probe schemes \cite{PhysRevSTAB.17.040702}— as well as the large energy chirp imparted by the THz field, which dominates over residual RF-induced jitter and nonlinear effects. The resulting suppression of the electron bunch arrival-time jitter leads to a significant reduction in shot-to-shot energy jitter after the LWFA.  Precise injection of a low-emittance bunch into a plasma accelerator results in minimal emittance growth; thus, the external injection of THz-controlled electron bunches successfully achieves the twin aims of producing stable, high-quality electron beams from an LWFA. 

The THz-driven beam control is studied analogously to conventional RF injectors, as discussed in Appendix~\ref{app:A}. By matching the linear energy chirp imprinted on the electron bunch ($h_1$) from the THz-driven waveguide with the first-order dispersion ($R_{56}$) of the magnetic arc, both the bunch length and arrival time can be precisely manipulated, as detailed in Appendix~\ref{app:B}. The voltage of THz-driven waveguides is currently limited by the availability of high-power THz sources; however, operating at high frequencies compensates for this constraint, allowing equivalent chirps to be imparted on electron bunches in a more compact setup compared to RF linacs.

To demonstrate the performance of the proposed concept, we performed two comprehensive numerical investigations of the generation and compression of electron bunches, and the injection and acceleration of these bunches in an LWFA. First, a start-to-end (S2E) simulation based on the CLARA/FEBE test facility at the STFC Daresbury Laboratory \cite{clara} was performed to validate the scheme under realistic accelerator conditions. The modeling accounted for space-charge effects, coherent synchrotron radiation (CSR), and potential arrival time jitter—coming from the injector. Second, we considered ideal Gaussian electron bunches to examine the fundamental capabilities of the proposed scheme and to identify potential pathways for improving the S2E model. The simulations integrated multiple codes: the start-to-end model of the electron injector is carried out using particle tracking codes: ASTRA \cite{astra}, GPT \cite{gpt}, and ELEGANT  \cite{Borland:2000gvh} to model the electron injector and beamline transport, an in-house code which was used for the THz-driven waveguide, and the quasi-3D particle-in-cell code \textsc{FBPIC} \cite{fbpic} for plasma dynamics.

Case 1: Numerical studies have already shown \cite{febe,febejit} that CLARA can deliver sub-50 fs electron bunches using its existing conventional RF infrastructure, suitable for a wide range of user applications. However, for stable production of high-energy, high-quality bunches after LWFA, the level of reported arrival time jitter remains a limiting factor. In this work, we implement our concept based on the facility and demonstrate how it can significantly enhance stability and unlock the full potential of both the facility and LWFA-based systems.
Initially, only RF-manipulated bunches without any THz control were injected into the plasma accelerator. These results were then compared to THz-controlled external injection to evaluate the effectiveness of the concept. Bunches from the injector delivered to THz-driven waveguides had a mean energy of $230.1\,$MeV, with the reference bunch having an energy spread of approximately $22.9\,\mathrm{keV}\,$rms, normalized emittance of $0.24\,\mu\mathrm{m}$ (both vertical and horizontal), transverse size $\simeq\,46.3\,\mu\mathrm{m}$ (vertical) and $\simeq\,61.0\,\mu\mathrm{m}$ (horizontal), bunch duration of $189.0\,\mathrm{fs}$ rms, and a charge of $5\,\mathrm{pC}$.

The DLW is a hollow rectangular waveguide \cite{Hibberd2020} with an interaction length of 20\,mm and operating frequency of 0.4\,THz, driven by an on-axis longitudinal THz field of 100\,MV/m \cite{thz_pulse}. The magnetic chicane, similar to the FEBE arc at the CLARA test-facility, consists of four dipoles, six quadrupole families, and two sextupoles, and can be tuned to provide an $R_{56}$ between $\pm 20\,$mm.

To prevent emittance growth inside the LWFA, the electron bunches are transversely matched into a plasma waveguide using quadrupoles, as described by the matching condition in Appendix~\ref{app:C} \cite{ROSSI201667,match}. The plasma is 9\,cm long with linear ramps at the entrance and exit of length 5.5 $\lambda_\beta$, where $\lambda_\beta=\frac{2\pi}{k_p}\sqrt{2\gamma}$, and $\gamma$ is the beam average Lorentz factor, and $k_p$ is the plasma wavenumber. The waveguide has a parabolic density profile of $n_p = n_{p,0} + r^2/\pi r_e \omega_0^4$, where $n_{p,0}$= 10$^{17}$ cm$^{-3}$ is the on-axis plasma density, $r$ is the radial coordinate, $r_e$ is the classical electron radius, and $\omega_0=\,$30$\,\mu$m is the laser waist (1/e$^2$ radius). Both the LWFA and THz pulse are driven by a 2.5\,J, 100\,TW laser with $\tau \simeq 25$\,fs pulse duration. 2\,J of the laser pulse energy is used to drive the LWFA.

By imprinting an energy chirp of $h_1 \sim 4.9\, \mathrm{MeV/ps}$ using the THz-driven waveguide, combined with $R_{56}\,=\,1.5\,\mathrm{cm}$ from the arc, the scheme intentionally increases the energy spread of the initial electron bunch. This configuration enables the longitudinal compression of the reference bunch from an initial $189.0\ \mathrm{fs\, rms}$ to $\sim\!7.8\ \mathrm{fs\, rms}$, resulting in a peak current of $\sim\!254.0\, \mathrm{A}$. Furthermore, this method effectively suppresses the arrival time jitter relative to the LWFA, reducing it from $98.4 \mathrm{\, } \pm \mathrm{\, } 9.8\ \mathrm{fs\, rms}$ to $8.0 \mathrm{\, } \pm \mathrm{\, } 0.8\ \mathrm{fs\, rms}$ as validated by start-to-end (S2E) modeling across 100 simulations (see Appendix~\ref{app:B} for further demonstration). 

\begin{figure}[h]
\includegraphics[width=0.5\textwidth]{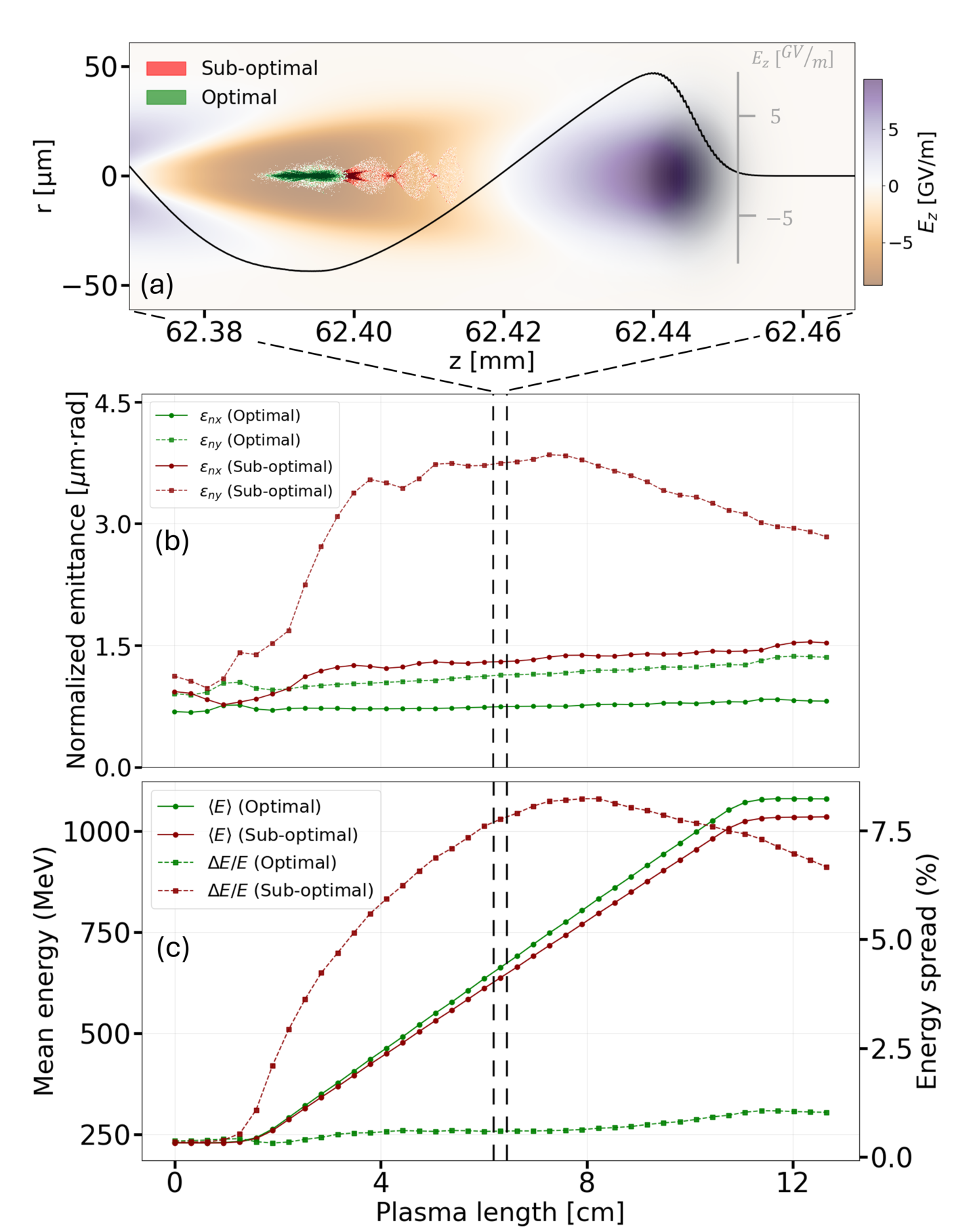}
\caption{Bunch parameter evolution along the accelerator for optimal and sub-optimal injection conditions. (a) Longitudinal electric field ($E_z$), shown in both 2D (color map) and 1D (line plot), taken at the midpoint of the plasma, with the corresponding positions of the optimal ($\mid$0.4$\mid\,\text{fs}$ temporal offset) and sub-optimal ($\mid$12.7$\mid\,\text{fs}$ temporal offset) bunches inside the wake structure. (b) Evolution of the normalized emittance along the plasma length, highlighting the severe transverse mismatch and subsequent $\epsilon_{ny}$ growth in the sub-optimal case. (c) Evolution of the mean energy ($E$, solid lines) and relative energy spread ($\Delta E/E\,$rms, dashed lines) along the plasma, demonstrating the necessity of precise timing to maximize energy gain and minimize energy spread ($1.0\%$ vs $6.6\%$).}
\label{fig:3} 
\end{figure}

Figure\,\ref{fig:3} presents the impact of optimal injection timing by comparing the bunch parameters for two different cases from the S2E model. The optimal bunch, defined by a time difference of only $\mid$0.4$\mid$\,fs relative to the reference bunch, is compared with a sub-optimal bunch that has a timing difference of $\mid$12.7$\mid$\,fs. Figure\,\ref{fig:3}(a) shows the longitudinal accelerating field at the midpoint of the plasma, together with the positions of the two bunches within the acceleration phase. Figure\,\ref{fig:3}(b) shows the evolution of the normalized emittance for both cases along the plasma length. The optimal bunch is matched into the plasma channel with normalized emittance of $0.8\,\mu\mathrm{m}$ (horizontal) and $0.9\,\mu\mathrm{m}$ (vertical), and transverse size of $3.5\,\mu\mathrm{m}$ (horizontal) and $4.1\,\mu\mathrm{m}$ (vertical), while the sub-optimal bunch is matched with normalized emittance of $1.1\,\mu\mathrm{m}$ (horizontal) and $1.5\,\mu\mathrm{m}$ (vertical), and transverse size of $5.8\,\mu\mathrm{m}$ (horizontal) and $8.2\,\mu\mathrm{m}$ (vertical). It is evident that the optimally injected bunch maintains its normalized emittance in both transverse directions, while the sub-optimal case shows a strong increase—predominantly in the $y$ direction—due to transverse mismatching and non-optimal injection.
Figure\,\ref{fig:3}(c) presents the mean energy and energy spread evolution for both cases. Both bunches are injected into the LWFA with a mean energy of approximately 229\,MeV and an initial energy spread of about 0.35\%, However, due primarily to the injection timing, the optimal bunch reaches 1079.7\,MeV with an energy spread of 1.0\%, while the sub-optimal one reaches 1035.4\,MeV with an energy spread of 6.6\%. This shows the importance of injection timing; only 12\,fs temporal offset produces a bunch of reduced energy and significantly increased energy spread and emittance.

\begin{figure}[h]
\includegraphics[width=0.5\textwidth]{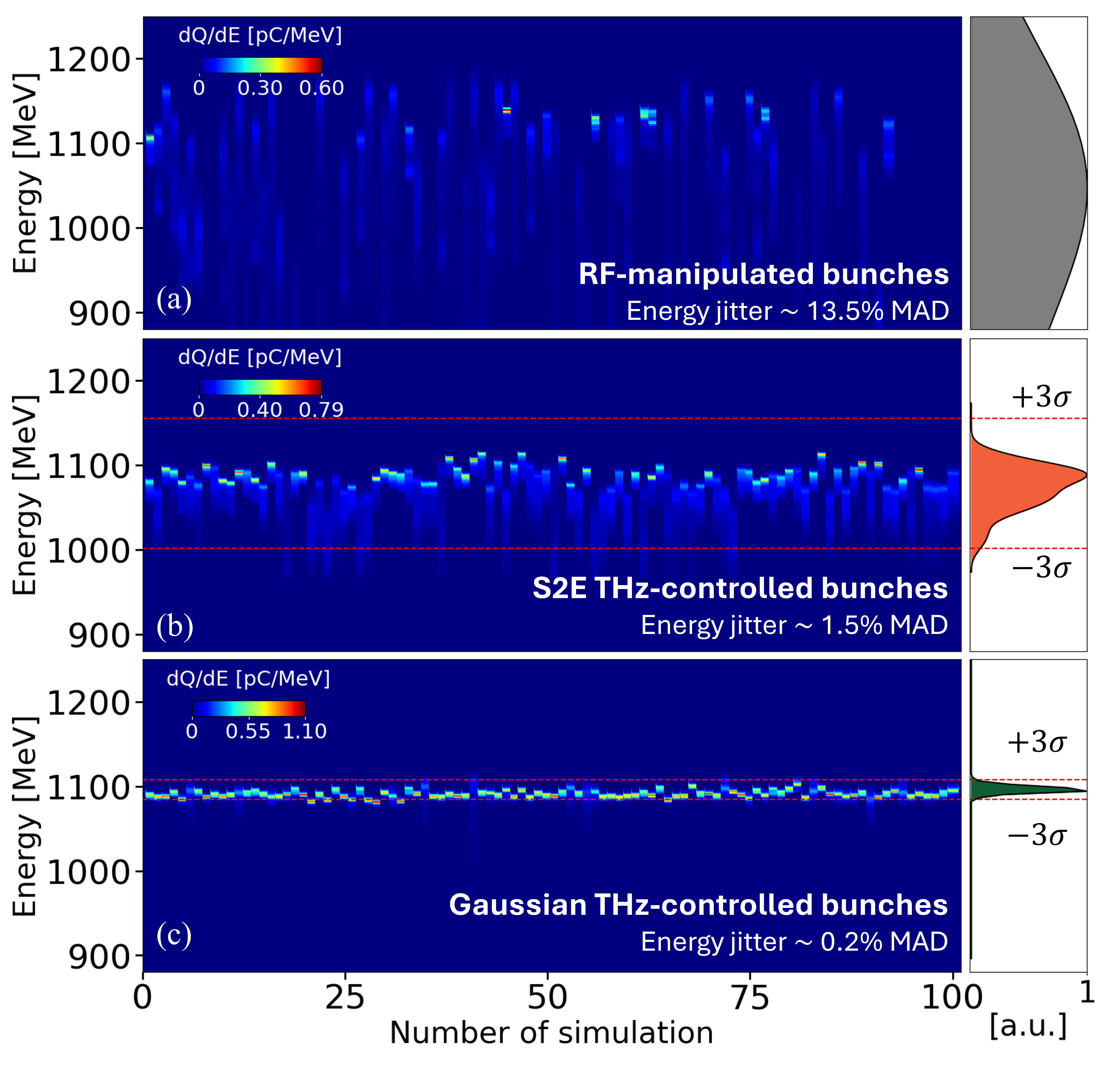}
    \caption{Shot-to-shot energy spectra (100 simulations) for (a) RF-manipulated, (b) S2E THz-controlled, and (c) Gaussian THz-controlled bunches. Left panels show spectral charge density ($dQ/dE$); right panels show integrated histograms with $\pm3\sigma$ bounds. THz control reduces energy jitter from 13.5\% MAD (RF-only) to 1.5\% (S2E) and 0.2\% (Gaussian) at GeV scales.}
    \label{fig:4}  
\end{figure}

To investigate the impact of the concept on the final beam quality, we modeled bunches with suppressed arrival time jitter that were injected on a shot-to-shot basis into the LWFA. For these analyses, the median of the energy spectrum and the corresponding median absolute deviation (MAD) are used to provide robust measures of the central energy of the beam and energy jitter, respectively, as commonly adopted in the laser–plasma community \cite{Winkler2025} (see supplementary materials for further details). For comparison, Fig.\ref{fig:4}(a) presents the energy spectra of RF-manipulated bunches (without THz control) after LWFA acceleration, showing a final energy of 1012.5 MeV median, and an energy jitter of 13.5\% MAD (20.0\% rms), primarily due to high arrival time jitter above 70\,fs\,rms. By implementing the THz-controlled scheme, Fig.\ref{fig:4}(b) shows that the electron beam achieves a final energy of 1079.5\,$\pm$\,16.4\,MeV (median $\pm$ MAD), nearly 1.5\% energy jitter (2.2\% rms), an energy spread of 1.1\% (median), a normalized emittance of 1.1\,$\pm$\,0.3\,$\mu$m (horizontal) and 1.4\,$\pm$\,0.2\,$\mu$m (vertical), and mean charge of 4.3\,$\pm$\,0.4\,pC. While only a small fraction of RF-manipulated shots produce high-quality bunches, the THz-controlled scheme achieves nearly an order of magnitude reduction in energy jitter, with 17\% of shots exhibiting an energy spread below 0.5\%. It should be noted that, with the current THz frequency of 0.4\,THz and the THz pulse energy, the electron bunches delivered by the injector—having bunch lengths above 180\,fs\,rms and arrival time jitter of 100\,fs\,rms\,—are not fully optimized for the given THz-driven waveguide. Therefore, a Gaussian electron bunch is considered to fully evaluate the capabilities of the proposed concept.

Case 2: The Gaussian bunch parameters in the second case are based on the ultimate expected machine development targets of CLARA \cite{febe}, representing an optimal performance expectation from the conventional elements of the scheme: 250\,MeV energy, 20 keV rms energy spread, $0.3\,\mu$m normalized emittance (both vertical and horizontal), transverse size $\sim\,30\,\mu$m (both vertical and horizontal), 5\,pC charge, and shorter bunches than the S2E model with a bunch length of 120\,fs rms. Using the same system (DLW, and arc) as employed for case 1, with rematched parameters, the reference bunch is compressed to $5.1\,\mathrm{fs}$\, with the initial arrival time jitter $100.1 \mathrm{\, } \pm \mathrm{\, } 10.1\ \mathrm{fs\, rms}$  suppressed to $3.4 \mathrm{\, } \pm \mathrm{\, } 0.3\ \mathrm{fs\, rms}$. After matching the beam into the plasma channel, Fig.\ref{fig:4}(c) shows the electron beam reaches a final energy of 1095.7\,$\pm$\,2.1\,MeV (median $\pm$ MAD), 0.2\% energy jitter (0.3\% rms), an energy spread of 0.2\% (median), a normalized emittance of 1.6\,$\pm$\,2.3\,$\mu$m (horizontal) and 0.5\,$\pm$\,0.1\,$\mu$m (vertical), and mean charge of 4.0\,$\pm$\,0.3\,pC. These results demonstrate a route to an order-of-magnitude reduction in energy spread compared to self-injection, alongside a significant suppression of energy jitter, establishing the proposed concept as a viable source of high-quality beams for GeV-level LWFA-based facilities.

In conclusion, the proposed concept of THz-controlled electron bunches for external injection into LWFA offers a promising route to high-charge, high-quality electron bunches with significantly reduced sensitivity to timing jitter and energy instabilities. Numerical results demonstrate that by taking advantage of temporal locking and the high-frequency nature of the THz field, the scheme achieves unprecedented jitter suppression—addressing a critical challenge even in advanced RF-based facilities. The stable, high-quality electron bunches generated through this method open a path toward multi-stage LWFA systems with transformative potential for applications such as compact free-electron lasers, direct injection into storage rings, and plasma-based colliders.

This work was supported by the United Kingdom Science and Technology Facilities Council (STFC) [Grant No. ST/V001612/1], and Dean's Doctoral Scholarship Award from the University of Manchester. We acknowledge the use of the high-performance cluster (CSF3) at the University of Manchester for the numerical simulations.


\bibliography{ref}
\onecolumngrid
\section{End Matter}

\twocolumngrid
\setcounter{secnumdepth}{2}
\appendix
\section{ Longitudinal Section Magnetic (LSM) modes in a rectangular DLW}
\label{app:A}
A set of orthonormal modes, known as longitudinal section magnetic (LSM$_{mn}$) modes, are supported in the rectangular DLW, where $m$ and $n$ denote the number of half-wavelength variations along the horizontal and vertical axes, respectively \cite{dlw}. The most appropriate mode for interaction with a beam is the LSM$_{11}$ mode, which provides an on-axis non-zero longitudinal electric field ($E_z$). If $v_p$ is the phase velocity of the THz wave and $v_e$ is the velocity of the electron bunch, when phase-velocity matching is achieved ($v_p = v_e$), the electron interacts synchronously with the LSM$_{11}$ mode of the THz field in a dielectric-lined waveguide.
\section{Bunch compression in an arc section}
\label{app:B}
After the DLW, the energy deviation of an electron at a longitudinal position $\zeta_0$, relative to a reference particle at $\zeta_{\mathrm{ref}}$ is given by $\delta = (\gamma_f - \gamma_{\mathrm{ref}})/\gamma_{\mathrm{ref}}$, where $\gamma_{\mathrm{ref}}$ is the Lorentz factor of the reference particle after DLW. Similarly to RF cavities, the energy deviation can be expanded as $\delta = \delta_0 + h_1 \zeta + h_2 \zeta^2 + \mathcal{O}(\zeta^3)$, where $\delta_0$ is the initial energy deviation of the single particle relative to the reference particle, mainly defined as uncorrelated energy spread. The coefficients $h_1$ and $h_2$ represent the linear and quadratic energy chirps, given by:
\begin{equation}
    h_1 = \frac{e V k \sin(\phi_{\mathrm{THz}})}{\gamma_{\mathrm{ref}}}, \quad
    h_2 = \frac{e V k^2 \cos(\phi_{\mathrm{THz}})}{2\gamma_{\mathrm{ref}}}.
    \label{Eq1}
\end{equation}

In Eq.\ref{Eq1}, $e$ is the elementary charge,  $V$ is the voltage inside the DLW, and, $k$ and $\phi_{THz}$ are the wavenumber and phase of the THz field, respectively.

A dispersive arc section transforms the longitudinal phase-space coordinates of a particle ($\zeta_0,\delta)$ to a final position as $ \zeta_f= \zeta_0 + R_{56}\delta+T_{566}\delta^2+O (\delta)$, where $R_{56}$, and $T_{566}$ are first and second-order longitudinal dispersive parameters of the arc, respectively. Using the energy deviation relation, the final rms bunch length of an electron bunch can be expressed as: 
 {\small
\begin{equation}
    \sigma_{\zeta_f} \approx \sqrt{R_{56}^2 \sigma_{\delta_0}^2 + \left[(1 + h_1 R_{56})^2 + 2 R_{56} \delta_0 (h_2 R_{56} + h_1^2 T_{566})\right] \sigma_{\zeta_i}^2}\, ,
\end{equation}
}where $\sigma_{\delta_0}$ is the rms of the initial uncorrelated energy spread of the bunch. Maximum compression requires matching condition ($R_{56}h_1=-1,\textbf{ }T_{566}{h_1^2}=h_2 R_{56}$). The compression mechanism and its impact on the longitudinal phase space are compared for both the S2E and Gaussian models in Fig.~\ref{fig:2}. By imprinting the necessary energy chirp (red) using the THz-driven waveguide and subsequently rotating the phase space in the arc transport (purple), the system achieves a compressed state from the initial injector stage (blue). This evolution is visualized in the longitudinal phase spaces for the S2E (c) and Gaussian (d) cases, where the normalized longitudinal spectral densities demonstrate a reference bunch length compression from $189.0\,\text{fs rms}$ to $7.8\,\text{fs rms}$ in the S2E model (a), and from $120.0\,\text{fs rms}$ to $5.1\,\text{fs rms}$ in the Gaussian model (b). Furthermore, the corresponding initial and final current profiles (e, f) illustrate the robustness of the arrival-time jitter suppression across both beam distributions; by effectively ``locking'' the reference bunch and the surrounding train to the LWFA-synchronized phase, the jitter is reduced to $8.0\,\text{fs rms}$ for the S2E model and $3.4\,\text{fs rms}$ for the Gaussian model.
\begin{figure}[h]
\includegraphics[width=0.5\textwidth]{fig2.png}
   \caption{THz-driven bunch compression and arrival-time jitter suppression. (a) Normalized longitudinal spectral density of the reference bunch illustrating compression from 189.0\,fs\,rms to 7.8\,fs\,rms for the S2E model, and (b) from 120.0\,fs\,rms to 5.1\,fs\,rms for the Gaussian model. (c, d) Longitudinal phase spaces of the reference bunch at successive stages — injector (blue), after the DLW (red), and after the arc (purple) — for the S2E and Gaussian cases, respectively. (e-f) Corresponding initial and final current profiles of the reference bunch (red line) and other bunches within the train (blue lines), showing suppression of the arrival-time jitter to 8.0\,fs\,rms and 3.4\,fs\,rms for the S2E and Gaussian models, respectively.}
    \label{fig:2}  
\end{figure}

\section{Transverse matching of electron bunches in a plasma waveguide}
\label{app:C}
To prevent emittance growth inside the LWFA, the electron bunches are transversely matched into a plasma waveguide using quadrupoles according to the matching condition:
\begin{equation}
    \sigma_{tr,match}=\sqrt[4]{\frac{2}{\gamma}}\sqrt{\frac{\epsilon_n}{k_p}},\quad
\end{equation}

where $\sigma_{tr,match}$ is the matched size in a plasma channel, $\epsilon_n$ is the normalized emittance, $\gamma$ is the beam average Lorentz factor, and $k_p$ is the plasma wavenumber.

\end{document}